\documentclass[10pt,a4paper]{article}
\usepackage{amsmath,amsthm,amsfonts,amssymb,fullpage,hyperref}
\usepackage[utf8]{inputenc}
\usepackage[T2A]{fontenc}
\numberwithin{equation}{section}

\newtheorem{proposition}{Proposition}
\newtheorem{theorem}{Theorem}
\allowdisplaybreaks
\def\i{{\rm i}}
\makeatletter
\def\section{\@startsection{section}{1}{\z@}%
            {-3.5ex \@plus -1ex \@minus -.2ex}%
            {2.3ex \@plus.2ex}%
            {\normalfont\large\bfseries}}
\def\subsection{\@startsection{subsection}{2}{\z@}%
            {-3.25ex\@plus -1ex \@minus -.2ex}%
            {1.5ex \@plus .2ex}%
            {\normalfont\normalsize \bfseries\itshape}}
\def\@seccntformat#1{\csname the#1\endcsname.~~}
\def\Appendix{\appendix
  \def\@seccntformat##1{Appendix~\csname the##1\endcsname.~~}}
\makeatother

\begin{document}
\newpage
\title{\Large Form factors of descendant operators in the Bullough-Dodd model}
\author{Oleg Alekseev\\[\medskipamount]
{\normalsize\it
Landau Institute for Theoretical Physics,}\\
{\normalsize\it 142432 Chernogolovka of Moscow Region, Russia}\\
{\normalsize\it and}\\
{\normalsize\it Center for Quantum Spacetime, Sogang University,}\\
{\normalsize\it Shisu-dong, Mapo-gu, Seoul 121-742 Korea}}
\date{}
\maketitle


\begin{abstract}
We propose a free field representation for the form factors of descendant operators in the Bullough-Dodd model. This construction is a particular modification of Lukyanov's technique for solving the form factors axioms. We prove that the number of proposed solutions in each level subspace of the chiral sectors coincide with the number of the corresponding descendant operators in the Lagrangian formalism. We check that these form factors possess the cluster factorization property. Besides, we propose an alternative free field representation which allows us to study analytic properties of the form factors effectively. In particular, we prove that the form factors satisfy non trivial identities known as the ``reflection relations''. We show the existence of the reflection invariant basis in the level subspaces for a generic values of the parameters.
\end{abstract}


\section{Introduction}

The form factors of local operators in two dimensional integrable models can be exactly obtained as the solutions to a system of functional equations known as the form factor axioms~\cite{KW,S,Sbook}. These equations specify the correct analytical properties of the form factors predefined by the scattering matrix of the model. In principle, one can solve these equations recursively. The space of solutions to the form factors axioms is supposed to give a faithful representation of the operator content of the model. However, the correct identification of the form factors of the specific operator within this space of solutions is in general non trivial problem. In the case of the exponential operators there exist two useful criteria which assist in the identification. The first one restricts the asymptotic behavior of the form factors at large values of rapidity~\cite{DM} while the second one is the so called cluster factorization property~\cite{DSC}. But the family of exponential operators are far from exhausting the full set of operators in the theory, which also contains descendant operators. The two criteria we discussed are not sufficient to solve the identification problem for these operators. In some cases the problem of identification was implemented. For example, in~\cite{Delfino_Niccoli:npb707} there appears the identification of the operator $T\bar T$ in the case of the Lee-Yang model and further extended to the sinh-Gordon model and to a series of perturbed minimal models in~\cite{Delfino_Niccoli:jhep05}. However, a general approach to solve this problem remains unknown.

Here we consider the Bullough-Dodd model~\cite{DB,ZS}. A convenient method for solving the form factor axioms is the free field representation proposed by Lukyanov~\cite{L}. By using this method the solutions to the form factor axioms in the Bullough-Dodd model that correspond to the exponential operators were found and completely identified~\cite{BL}. Following the guidelines of~\cite{FL} we find the free field representation for the form factors of descendant operators.

The proposed free field representation allows us to study analytic properties of the form factors efficiently. In particular, we prove that the form factors satisfy the reflection relations~\cite{FLZZ}. These relations originate from the similar relations in the Liouville theory and establish connection between the exponential operators~$e^{a\varphi}$ with different values of the parameter $a$~\cite{ZZreflection}. Besides, we prove the existence of the reflection invariant bases in the Fock spaces.

This paper is organized as follows. In Section~\ref{BD} we describe the Bullough-Dodd model and specify its operator contents. In Section~\ref{FF} we briefly describe Lukyanov's free field representation for the form factors of exponential operators. We introduce auxiliary construction and find the free field representation for the set of solutions to the form factor axioms. The number of these solutions on each level subspace of the Fock module is shown to be coincide with the dimension of the corresponding level subspace in the Lagrangian formalism. We prove that these solutions satisfy the cluster factorization property and study their asymptotic behavior. In Section~\ref{RR} we introduce alternative free field representation which is used to obtain recurrence relations between the form factors and prove the reflection properties. In Section~\ref{E} we consider an explicit example, namely, level $(2,0)$ descendant operators. Finally, we draw our conclusions in Section~\ref{C}.


\section{The Bullough Dodd model}\label{BD}

The Bullough-Dodd model is defined by the Eucledean action~\cite{DB,ZS}
\begin{equation}
	S_{BD}=\int d^2x\left(\frac{1}{16\pi}(\partial_\nu\varphi)^2+\mu(e^{\sqrt2b\varphi}+2e^{-\frac{b}{\sqrt2}\varphi})\right).
\end{equation}
Here $b$ is the coupling constant and $\mu$ is the regularized mass parameter. The spectrum of the model consist of the single particle $A$ of mass $M$ which can be expressed in terms of the parameter $\mu$~\cite{FLZZ}. This particle appears as the bound state of itself in the scattering processes
\begin{equation}\label{BD:AA}
	A\times A\to A\to A\times A.
\end{equation}
Hereinafter we shall use the following notation
\begin{equation}\label{BD:Q}
	Q=b+b^{-1},\quad \omega=e^{\i\pi/3}.
\end{equation}
Also, we shall use the light-cone coordinates
\begin{equation}\nonumber
	z=x^1-x^0,\quad \bar z=x^1+x^0,\quad \partial=\frac{\partial}{\partial z},\quad \bar \partial=\frac{\partial}{\partial \bar z}.
\end{equation}
The integrability of the model implies that the scattering processes are purely elastic~\cite{ZZ}. Therefore, the $n$-particle $S$-matrix factorizes into the $n(n-1)/2$ two-particle scattering amplitudes~\cite{FZ}
\begin{equation}\label{BD:S}
	S(\theta)=\frac{\tanh\frac{1}{2}(\theta+\frac{2\i\pi}{3})\tanh \frac{1}{2}(\theta-\frac{2\i\pi}{3bQ}) \tanh\frac{1}{2}(\theta-\frac{2\i\pi b}{3Q})}{\tanh\frac{1}{2}(\theta-\frac{2\i\pi}{3}) \tanh\frac{1}{2}(\theta+\frac{2\i\pi}{3Qb})\tanh\frac{1}{2}(\theta+\frac{2\i\pi b}{3Q})}.
\end{equation}
The $S$-matrix is invariant under the weak-strong coupling constant duality $b\to b^{-1}$. A simple pole of the scattering matrix located at $\theta=2\pi \i/3$ corresponds to the bound state of the particle $A$ itself in the scattering process~\eqref{BD:AA}. 

Let us consider the operator contents of the model. The space of local operators of the Bullogh-Dodd model consists of the exponential operators
\begin{equation}
	V_a(x)=e^{a \varphi(x)}
\end{equation}
and their descendants, i.e.\ the linear combinations of the fields
\begin{equation}\label{BD:desc}
\partial^{n_1}\varphi\ldots\partial^{n_r}\varphi\bar{\partial}^{\bar n_1}\varphi\ldots\bar{\partial}^{\bar n_s}\varphi \, e^{a\varphi(x)}.
\end{equation}
Here the pair of integers $(n,\bar n)$ given by
\begin{equation}
	n=\sum_{i=1}^r n_i,\quad \bar n=\sum_{j=1}^s \bar n_j
\end{equation}
is called the level of descendant operator. The numbers $n$ and $\bar n$ separately are called chiral and antichiral level correspondingly. Any exponential operator is characterized by its scaling dimension $\Delta_a$ in the ultraviolet regime which is related to the parameter $a$. The scaling dimension of the corresponding descendant operator at the level $(n,\bar n)$ is given by $\Delta_a+n+\bar n$, while its spin is $n-\bar n$. 

Let us consider the radial quantization picture at some point in the Eucledean plane, e.g.\ $x=0$. In the vicinity of this point the field $\varphi(x)$ can be expanded as
\begin{equation}
	\varphi(x)=\mathbf Q-\i\mathbf P\log z\bar z+\sum_{m\neq0}\frac{\mathbf a_m}{\i m}z^{-m}+\sum_{m\neq0}\frac{\bar{\mathbf a}_m}{\i m}\bar z^{-m}.
\end{equation}
Here the operators $\mathbf Q$, $\mathbf P$, $\mathbf a_m$ and $\bar{\mathbf a}_m$ form a Heisenberg algebra with the following commutation relations
\begin{equation}\label{BD:alg}
	[\mathbf P,\mathbf Q]=-\i,\quad[\mathbf a_m,\mathbf a_n]=m\delta_{m+n,0},\quad [\bar{\mathbf a}_m,\bar{\mathbf a_n}]=m\delta_{m+n,0}.
\end{equation}
In the radial quantization picture the exponential operator $V_a(0)$ corresponds to the highest weight vector $|a\rangle_{rad}$ defined by the relations
\begin{equation}
	\mathbf a_m|a\rangle_{rad}=\bar{\mathbf a}_m|a\rangle_{rad}=0\quad(m>0),\quad \mathbf P|a\rangle_{rad}=a|a\rangle_{rad},\quad |a\rangle_{rad}=e^{a\mathbf Q}|vac\rangle_{rad}.
\end{equation}
Let $\mathcal F_a$ be the Fock module spanned on the vectors generated by the elements $\mathbf a_{-m}$ $(m<0)$ acting on the highest weight vector $|a\rangle_{rad}$. Similarly, the $\bar{\mathcal F_a}$ is a Fock module spanned on the vectors generated by the elements $\bar{\mathbf a}_{-m}$ $(m<0)$ acting on the same highest weight vector. Evidently, the modules $\mathcal F$ and $\bar{\mathcal F}$ are isomorphic. The space of states is given by the tensor product $\mathcal F_a\otimes\bar{\mathcal F}_a$ of chiral and antichiral components. Up to some numerical factor the descendant operators~\eqref{BD:desc} correspond to the vectors
\begin{equation}\label{BD:vec}
\mathbf a_{-n_1}\ldots\mathbf a_{-n_r}\bar{\mathbf a}_{-\bar n_1}\ldots\bar{\mathbf a}_{-\bar n_s}|a\rangle_{rad}\quad(0<n_1\leq\ldots\leq n_r,\ 0<\bar n_1\leq\ldots\leq \bar n_s).
\end{equation}
Each module $\mathcal F_a$ admits a natural gradation into the level subspaces $\mathcal F_a=\oplus_{n=0}^{\infty}\mathcal F_{a,n}$, where each level subspace $\mathcal F_{a,n}$ is spanned on the vectors~\eqref{BD:vec} with $\bar n=0$ and $\sum n_i=n$. The dimensions of these subspaces are given by the following generating function
\begin{equation}
\sum_{n=0}^\infty q^n\dim\mathcal F_{a,n}=\prod_{m=1}^\infty\frac{1}{1-q^m}.
\end{equation}
The dimension of the level $(n,\bar n)$ subspace if given by the product $\dim\mathcal F_{a,n}\cdot\dim\bar{\mathcal F}_{a,\bar n}$.


\section{Free Field representation for form factors}\label{FF}


\subsection{Form factors of exponential operators}

Let us briefly describe Lukyanov's free field representation for the form factors of exponential operators in the Bullough-Dodd model~\cite{BL}. First, we consider a pair of operators $\Lambda^+(\theta)$, $\Lambda^-(\theta)$ and define the one and two-point trace functions as follows
\begin{equation}\label{FFE:LL}
	\langle\langle\Lambda^\sigma(\theta)\rangle\rangle=1,\quad
	\langle\langle \Lambda^{\sigma'}(\theta')\Lambda^\sigma(\theta)\rangle\rangle=\left [R(\theta-\theta')\right]^{\sigma'\sigma},\quad \sigma',\sigma=\pm,
\end{equation}
where the function $R(\theta)$ is the two-point minimal form-factor given by~\cite{FMS}
\begin{equation}
	R(\theta)=\exp\Bigl(-4\int_0^\infty\frac{dt}{t}\frac{\cosh \frac{t}{6}\sinh\frac{tb}{3Q}\sinh\frac{t}{3Qb}}{\sinh t\sinh\frac{t}{2}}\cosh(t-\frac{\i\theta t}{\pi})\Bigr).
\end{equation}
The multi-point functions can be calculated by means of the Wick's theorem. Let us define the normal ordering procedure $:\ldots:$ by the following relation
\begin{equation}
	\Lambda^{\sigma_N}(\theta_N)\ldots \Lambda^{\sigma_1}(\theta_1)=:\Lambda^{\sigma_N}(\theta_N)\ldots \Lambda^{\sigma_1}(\theta_1):\prod_{1\leq i<j\leq N}\langle\langle \Lambda^{\sigma_j}(\theta_j)\Lambda^{\sigma_i}(\theta_i)\rangle\rangle.
\end{equation}
Hereinafter we shall use the notation
\begin{equation}
	\Lambda^0(\theta)=:\Lambda^+(\theta-\frac{\i\pi}{3})\Lambda^-(\theta+\frac{\i\pi}{3}):.
\end{equation}
The Lukyanov's generators are given by
\begin{equation}\label{FFE:T}
	T(\theta)=\rho\, (e^{\i \pi p}\,\Lambda^+(\theta)+e^{-\i\pi p}\,\Lambda^-(\theta)+ h\, \Lambda^0(\theta)),
\end{equation}
where we introduce the follwing notation
\begin{equation}\label{FFE:const}
	\begin{aligned}
	h&=2\sin\frac{\pi(b-b^{-1})}{6Q},\\
	p&=\frac{4\sqrt{2}a-b+b^{-1}}{6Q}-\frac{1}{2},\\
	\rho&=\sqrt{\frac{\sin\frac{\pi}{3}}{\sin\frac{2\pi b}{3Q}\sin\frac{2\pi}{3Qb}}}\exp\Bigl(2\int_0^\infty\frac{dt}{t}\frac{\cosh\frac{t}{6}\sinh\frac{tb}{3Q}\sinh\frac{t}{3Qb}}{ \sinh t\cosh\frac{t}{2}}\Bigr).
	\end{aligned}
\end{equation}
The form factors of exponential operators are given by the multi-point trace functions, namely
\begin{equation}\label{FFE:F}
	\langle vac|V_a(0)|\theta_1,\ldots,\theta_N\rangle\equiv \langle e^{a\varphi}\rangle f_{a}(\theta_1,\ldots,\theta_N)=\langle e^{a\varphi}\rangle\langle\langle T(\theta_N) \ldots T(\theta_1)\rangle\rangle,
\end{equation}
where $\langle e^{a\varphi}\rangle$ is the vacuum expectation value of the corresponding exponential operator~\cite{FLZZ}. The functions $f_{a}(\theta_1,\ldots,\theta_N)$ are analytic functions in the variables $\theta_i$ with complicated analytic structure. Taking into account~\eqref{FFE:LL} we immediately conclude that these functions can be represented as follows
\begin{equation}
	f_{a}(\theta_1,\ldots,\theta_N)=\rho^N J_{N,a}(e^{\theta_1}\ldots,e^{\theta_N})\prod_{i\leq i<j\leq N}R(\theta_i-\theta_j).
\end{equation}
Here the functions $J_{N,a}(x_1,\ldots,x_N)$ are symmetric rational functions in the variables $x_i=e^{\theta_i}$. Their pole structure is determined by the form factor axioms. The only poles located at relative rapidity difference $\theta_{ij}=i\pi$ and $\theta_{ij}=2\pi i/3$ will be referred to as the kinematical and the bound state poles correspondingly.


\subsection{Form factors of descendant operators}\label{FFD}

We argue that the form factors of descendant operators possess the free field representation that is very similar to Lukyanov's one~\eqref{FFE:F}. The main idea is to modify the generators $T(\theta)$. Let us consider a commutative algebra $\mathcal A=\oplus_{n=0}^\infty\mathcal A_n$ generated by the elements $\lbrace \alpha_{-n}\rbrace$ with $n>0$. Each level subspace $\mathcal A_n$ is spanned on the elements $\prod \alpha_{-n_i}$ such that $\sum n_i=n$. We also introduce another copy $\bar{\mathcal A}$ of the algebra $\mathcal A$ generated by the elements $\lbrace\bar \alpha_{-n}\rbrace$. The canonical homomorphism between these algebras is defined by the following relation: for any $h\in\mathcal A$ let us define $\bar h\in\bar{\mathcal A}$ according to the rule $\alpha_{-n}\to\bar \alpha_{-n}$. The element $g=h\bar h'$ will be referred to as the level $(n,\bar n)$ descendant if $h\in \mathcal A_n$ and $\bar h'\in\bar{\mathcal A}_{\bar n}$.

Let us define a bracket on the algebra $\mathcal A$ by
\begin{equation}\label{FFD:bracket}
	\left(\prod_{m=1}^\infty \alpha_{-m}^{k_m},\prod_{m=1}^\infty \alpha_{-m}^{l_m}\right)=\prod_{m=1}^\infty k_m!\delta_{k_m,l_m}
\end{equation}
and consider the currents
\begin{equation}\label{FFD:ab}
	\begin{aligned}
	a(z)&=\exp\bigg(\sum_{m=1}^\infty \alpha_{-m}z^m\bigg),\\
	b(z)&=\exp\bigg(-\sum_{m=1}^\infty \alpha_{-m}(-z)^m\bigg),\\
	c(z)&=\exp\bigg(\sum_{m=1}^\infty (\omega^{-m}-(-1)^m \omega^m)\alpha_{-m}z^m\bigg).
	\end{aligned}
\end{equation}
By using these currents we modify the Lukyanov's generators as follows
\begin{equation}
	\mathcal T(\theta)=\rho\big(e^{\i\pi p}\, a(e^\theta)\bar b(e^{-\theta})\Lambda^+(\theta)+e^{-\i\pi p}\, b(e^\theta)\bar a (e^{-\theta})\Lambda^-(\theta)+h\, c(e^\theta)\bar c(e^{-\theta})\Lambda^0(\theta)\big).
\end{equation}
For any element $g\in\mathcal A\otimes\bar{\mathcal A}$ consider the function
\begin{equation}\label{FFD:f}
	f^g_a(\theta_1,\ldots,\theta_N)=(\langle\langle \mathcal T(\theta_N),\ldots,\mathcal T(\theta_1)\rangle\rangle,g).
\end{equation}
It is straightforward to check that this function is a solution to the form factor axioms. Indeed, the Watson's theorem and the crossing symmetry condition are evidently satisfied while the kinematical and the bound state pole conditions are satisfied if the currents~\eqref{FFD:ab} are subjected to the following conditions
\begin{equation}\nonumber
	a(z)b(-z)=1,\quad c(z) = a(z \omega^{-1})b(z \omega),
\end{equation}
which are the case. Consequently, the function~\eqref{FFD:f} determines the form factors of an operator from the Fock space $(\mathcal F\otimes\bar{\mathcal F})V_a(x)$. The descendant operator which corresponds to this function will be denoted by $V_a^g(x)$ that is
\begin{equation}\label{FFD:V}
	\langle vac|V^g_a(x)|\theta_1,\ldots,\theta_N\rangle=\langle e^{a\varphi}\rangle f_a^g(\theta_1,\ldots,\theta_N).
\end{equation}
From expression~\eqref{FFD:f} it follows that the solutions $f_a^g$ to the form factor axioms can be represented as follows
\begin{equation}\label{FFD:fJ}
	f_{a}^g(\theta_1,\ldots,\theta_N)=\rho^N J_{N,a}^g(e^{\theta_1},\ldots,e^{\theta_N})\prod_{i<j}^NR(\theta_i-\theta_j).
\end{equation}
Here the functions $J_{N,a}^g(x_1,\ldots,x_N)$ are symmetric rational functions in the variables $x_i=e^{\theta_i}$ with the kinematical and bound state poles located at relative rapidity differences $\theta_{ij}=i\pi$ and $\theta_{ij}=\pm2i\pi/3$ correspondingly. Evidently, the $J_{N,a}(x_1,\ldots,x_N)$ functions which correspond to the exponential operators are given by $J^1_{N,a}(x_1,\ldots,x_N)$. It is straightforward to get an explicit expression for the function $J_{N,a}^g(x_1,\ldots,x_N)$, namely
\begin{multline}\label{FFD:Jg}
	J_{N,a}^g(x_1,\ldots x_N)=\sum_{I_++I_-+I_0=I}h^{\#I_0}e^{(\#I_+-\#I_-)\i\pi p}\, P^g(X_+|X_-|X_0)\times\\
	\times\prod_{i\in I_+,j\in I_-,k\in I_0}f\Bigl(\frac{x_i}{x_j}\omega\Bigr)f\Bigl(\frac{x_i}{x_j}\omega^2\Bigr)f\Bigl(\frac{x_i}{x_k}\omega\Bigr)f\Bigl(\frac{x_j}{x_k}\omega^{-1}\Bigr)\prod_{(p<q)\in I_0}f\Bigl(\frac{x_p}{x_q}\Bigr).
\end{multline}
where
\begin{equation}\label{FFD:fx}
	f(x)=1+\frac{h^2-1}{x+x^{-1}-1}.
\end{equation}
In expression~\eqref{FFD:Jg} we introduce a set of integers $I=\lbrace1,\ldots,N\rbrace$ and the sum is taken over all decompositions of the set $I$ into the three subsets $I_\sigma$, $\sigma=\lbrace+,-,0\rbrace$, such that $I_+\cup I_-\cup I_0=I$ and $I_{\sigma'}\cap I_\sigma=\varnothing$ if $\sigma'\neq\sigma$. With every subset $I_\sigma$ we associate the subset $X_\sigma=\lbrace x_i|i\in I_\sigma\rbrace$. The functions $P^g(X|Y|Z)$ are polynomials defined by the following relations
\begin{equation}
	\begin{aligned}\label{FFD:P}
	&P^{\alpha_{-m}}(X|Y|Z)= S_m(X)-(-1)^m S_m(Y)+(\omega^{-m}-(-1)^m\omega^m)S_m(Z),\\
	&P^{\bar \alpha_{-m}}(X|Y|Z)=S_{-m}(Y)-(-1)^m S_{-m}(X)+(\omega^{-m}-(-1)^m\omega^m)S_{-m}(Z),\\
	&P^{g_1g_2}=P^{g_1}P^{g_2},\quad P^{c_1g_1+c_2g_2}=c_1P^{g_1}+c_2P^{g_2}\quad (\forall g_1,g_2\in\mathcal A\otimes\bar{\mathcal A},\ c_1,c_2\in\mathbb C).
	\end{aligned}
\end{equation}
Here we denote the power sums of order $m$ by
\begin{equation}
	S_m(x_1,\ldots,x_N)=\sum_{i=1}^N x^m.
\end{equation}

In summary, in the framework of the free field representation we obtain the set of solutions $f_a^g$ to the form factor axioms. Besides, in~\eqref{FFD:fJ} and~\eqref{FFD:Jg} we give explicit expressions for these functions. The form factors of exponential operators which correspond to the case $g=1$ have been studied in detail~\cite{BL}. We want to prove that the space of proposed solutions can be bijectively mapped into the space of descendant operators in the Lagrangian formalism. In other words, we want to prove that for any descendant operator in the Lagrangian formalism there exist a solution to the form factor axioms given by certain linear combination of the functions $f_a^g$.

It is a challenging problem to make an identification between the states from the space $\mathcal A\otimes\bar{\mathcal A}$ and those ones from $\mathcal F\otimes \bar{\mathcal F}$. Certain requirements are necessary to identify the form factors of a specific operator among all these functions. In the case of exponential operators the solution to this problem is known. It is sufficient to impose restriction on the asymptotic behavior of the form factors at large values of rapidities~\cite{DM} and demand the cluster factorization property to be satisfied~\cite{DSC}. However, additional criteria are required in the case of descendant operators. This problem is not solved in general cases. Nevertheless, let us consider the analytic properties of the proposed set of functions.


\subsection{Cluster factorization property and asymptotic behavior}

We consider the element $g=h\bar h'$ such that $h\in\mathcal A$, $\bar h'\in\bar{\mathcal A}$. We are interested in the asymptotic behavior of the function which corresponds to this element, namely
\begin{equation}
	f^{h\bar h'}_a(\theta_1,\ldots,\theta_n,\theta_{n+1}+\Lambda,\theta_N+\Lambda),
\end{equation}
as $\Lambda\to\infty$. It is convenient to study the asymptotic of this function using the representation~\eqref{FFD:fJ}. Notice that $R(\theta\pm \Lambda)\to1$ as $\Lambda\to\infty$. Besides, is straightforward to check that in this limit we have
\begin{equation}
	P^{h\bar h'}(Xe^\Lambda,X'|Ye^\Lambda,Y'|Z e^\Lambda,Z')\simeq P^h(X e^\Lambda|Y e^\Lambda|Z e^\Lambda)P^{\bar h'}(X'|Y'|Z').
\end{equation}
Therefore, we get the cluster factorization property for the functions $f_a^{h\bar h'}$, namely
\begin{equation}\label{FP:f}
	f^{h\bar h'}_a(\theta_1,\ldots,\theta_n,\theta_{n+1}+\Lambda,\ldots,\theta_N+\Lambda)\simeq f^{h}_a(\theta_{n+1}+\Lambda,\ldots,\theta_N+\Lambda)f^{\bar h'}_a(\theta_1,\ldots,\theta_n),
\end{equation}
as $\Lambda\to+\infty$. The cluster factorization property for the form factors of the descendant operators immediately follows from~\eqref{FFD:V}. The form factors of the level $(n,\bar n)$ descendant operators factorizes into the form factors of the chiral level $n$ and antichiral level $\bar n$ descendant operators. The cluster factorization property has been argued in~\cite{Delfino_Niccoli:npb707} and has been proven for the form factors of $(n,\bar n)$ descendant operators with $n,\bar n<8$ for the Lee-Yang model~\cite{DN}.

Let us discuss certain consequences which follow from the cluster factorization property. We consider the chiral elements $h\in\mathcal A_n$ and $\bar h'\in\bar{\mathcal A}_{\bar n}$. The corresponding descendant operators will be denoted by $V_a^h$ and $V_a^{\bar h'}$. The scaling dimensions of these operators in the ultraviolet limit are given by $\Delta_a^{h}=\Delta_a+n$ and $\Delta_a^{\bar h'}=\Delta_a+\bar n$, where $\Delta_a$ is the scaling dimension of the  corresponding exponential operator. The spins of these descendant operators are the following, $s^h=n$ and $s^{\bar h'}=-\bar n$. From the cluster factorization property we immediately conclude that the element $h$ corresponds to the chiral level $n$ descendant operators while the element ${\bar h'}$ corresponds to antichiral level $\bar n$ descendant operators.

Let us consider the generic descendant operator $V_a^g$ related to the element $g=h\bar h'$. The scaling dimension of this operator is given by $\Delta^{h\bar h'}_a=\Delta_a+n+\bar n'$ while its spin is the following, $s^{h\bar h'}=n-\bar n$. From the cluster factorization property we conclude that this operators is given by certain linear combination of the level $(l,\bar l)$ descendants where $l\leq n$ and $\bar l\leq \bar n$. Therefore, that the cluster factorization property do not impose rigorous conditions on the descendant operators which contribute to a particular function~$f_a^g$.


\subsection{Descendants counting}

In this subsection we shall prove that the number of independent solutions $f_a^g$ to the form factor axioms in each level subspace coincide with the number of the descendant operators in the corresponding level subspace in the Lagrangian formalism. First, let us prove the following theorem
\begin{theorem}\label{FFD:dct}
For generic values of the parameter $a$ the map~\eqref{FFD:f} from the algebra $\mathcal A\otimes \bar{\mathcal A}$ into the space of the functions $f_a^g$ is a bijection.
\end{theorem}
Let us prove that the different elements $g$ correspond to the different functions $f_a^g$. First, we consider the chiral element $g=h\in\mathcal A$. Let us prove the linear independence of the set of polynomials~\eqref{FFD:P}. Indeed, for large enough $N$ the functions
\begin{equation}
	z_m=S_m(X)-(-1)^mS_m(Y)+(\omega^{-m}-(-1)^m \omega^m)S_m(Z)
\end{equation}
are functionally independent. Therefore, the linear independence of the set of polynomials $P^h(X|Y|Z)$ reduces to the evident linear independence of the monomials $z_1^{k_1}\ldots z_s^{k_2}$.

Now, let us consider the asymptotic of the function $J_{N,a}^h$ as the parameter $a\to-\i\infty$. It is easy to check that
\begin{equation}\label{DC:as}
	e^{-\frac{\i\pi(4\sqrt2a-b+1/b)}{6Q}+\frac{\i\pi}{2}}J_{N,a}^h(x_1,\ldots,x_N)\Big|_{a\to -\i\infty}=(a(x_1)\ldots a(x_N),h)=P^h(X|\varnothing|\varnothing).
\end{equation}
This expression defines a map from the algebra $\mathcal A$ to the algebra of polynomials in the variables $z_i=P^h(X|\varnothing|\varnothing)$. This map is evidently invertible. The linear independence of the polynomials $P^h(X|Y|Z)$ was already proved. Consequently, different elements from the algebra $\mathcal A$ correspond to different functions $f_a^h$. Now we apply the deformation argument. Since the map from the space of elements $h\in \mathcal A$ to the space of form factors $f_a^h$ is a bijection at one point in the parameter $a$ and the form factors are analytic functions in this parameter this map is a bijections for nearly all values of the parameter $a$. Hence, we prove that the form factors $f_a^h$ with different values of $h$ differs. Besides, one can easily prove that $\dim\mathcal A_n=\dim\mathcal F_n$.

Now let us consider the generic element $g=h\bar h'\in\mathcal A\otimes\bar{\mathcal A}$. Taking into account the cluster factorization property we conclude that the form factors of the different elements differs only if chiral and antichiral components of these form factors differs. Hence, we proved Theorem~\ref{FFD:dct}.

As an immediate consequence of the theorem we have
\begin{proposition}\label{dim}
For generic values of the parameter $a$ the dimension of the space of the operators $V_a^g$ with $g\in\mathcal A_n\otimes \bar{\mathcal A}_{\bar n}$ is equal to the dimension of the corresponding subspace of the Fock space $\dim(\mathcal F_n\otimes\bar{\mathcal F}_{\bar n})$.
\end{proposition}
Note that in~\cite{Delfino_Niccoli:npb799} it has been proven that the space of local operators in the Lee-Yang model is isomorphic to that of the corresponding ultraviolet CFT.

\subsection{Integrals of motion}

The Bullough-Dodd model possesses a set of commuting integrals of motion $I_s$ of odd integer spin $s$ except the multipliers of $3$, i.e. 
\begin{equation}\label{IM:s}
	s=6n\pm1\quad n=0,1,2\ldots
\end{equation}
The local integrals of motion are diagonalized by the asymptotic states and the corresponding eigenvalues are given by
\begin{equation}\label{IM:I}
	I_s|\theta_1,\ldots,\theta_N\rangle=\sum_{i=1}^N e^{s\theta_i}|\theta_1,\ldots,\theta_N\rangle,
\end{equation}
where the proper normalization of the integrals of motion is assumed. On the other hand, let us consider the chiral element $\alpha_{-m}$. This element produces a common factor in all terms in~\eqref{FFD:Jg}, provided that the following relations are satisfied
\begin{equation}
	e^{\i\pi m}=-1,\quad e^{-\frac{\i\pi}{3}m}+e^{\frac{\i\pi}{3}m}=1.
\end{equation}
The first relation  shows that the value of $m$ need to be odd, while the second one shows that the allowed values $m$ are those of~\eqref{IM:s}. For the corresponding elements $\alpha_{-s}$ the common factor in all terms in~\eqref{FFD:Jg} is of the form
\begin{equation}
	P^{\alpha_{-s}}(X_+|X_-|X_0)=S_s(X_+)+S_s(X_-)+S_s(X_0)=S_s(X),
\end{equation}
where $X=X_-\cup X_+\cup X_0$. Consequently, we get
\begin{equation}\label{IM:f}
	f^{\alpha_{-s} g}_a(\theta_1,\ldots,\theta_N)=\sum_{m=1}^N e^{s \theta_m}f_a^g(\theta_1,\ldots,\theta_N),
\end{equation}
for any $g\in\mathcal A\otimes \bar{\mathcal A}$. Comparing~\eqref{IM:I} with~\eqref{IM:f} we conclude that the elements $\alpha_{-s}$ correspond to the appropriately normalized spin $s$ integrals of motion $I_s$ while the antichiral element $\bar \alpha_s$ correspond to the integral of motion $I_{-s}$, i.e.
\begin{equation}
	V^{\alpha_{-s}g}_a(x)=[V^g_a(x),I_s],\quad V^{\bar\alpha_{-s}g}(x)=[V^g_a(x),I_{-s}].
\end{equation}


\section{Recurrent relations and reflection property for descendant operators}\label{RR}

In this section we shall prove that the form factors of descendant operators satisfy the reflection relations. These relations establish connection between the exponential operators with different values of the parameter $a$. Namely, up to some $a$-dependent factor the following operators are supposed to be coincide
\begin{equation}\label{RR:Vdef}
	V_a(x)=R(a)\ V_{Q_L-a}(x),\quad V_{-a}(x)=R'(a)\ V_{-Q'_L+a}(x),
\end{equation}
where we introduce the notation
\begin{equation}\label{RR:Q}
	Q_L=\frac{1}{\sqrt2b}+\sqrt2b,\quad Q'_L=\frac{\sqrt2}{b}+\frac{b}{\sqrt2}.
\end{equation}
The functions $R(a)$ and $R'(a)$ are the reflection amplitudes~\cite{FLZZ}. These reflection relations originates from the similar relations for exponential operators in the Liouville theory~\cite{ZZreflection}. Indeed, the Bullough-Dodd model can be treated as two different perturbed Liouville theories with charges $Q_L$ or $Q'_L$ depending on which exponent in~\eqref{BD:S} is taken as the perturbing operator. By using recursion relations one can prove that the $J$ functions of the exponential operators possess the following reflection properties~\cite{A}
\begin{equation}
	J_{N,a}(x_1,\ldots,x_N)=J_{N,Q_L-a}(x_1,\ldots,x_N),\quad J_{N,-a}(x_1,\ldots,x_N)=J_{-Q'_L+a}(x_1,\ldots,x_N).
\end{equation}
We argue that the correspondence between the exponential operators~\eqref{RR:Vdef} can be extended to the whole space of descendant operators. More specifically, for any descendant operator of the exponential field $V_a^g$ there exist a descendant operator of the exponential field $V_{wa}^{g'}$ such that the form factors of these operators coincide. Here we introduce the notation $w$ for the element of the finite group $\mathcal W$ generated by the elements $w_1$ and $w_2$ such that
\begin{equation}
	w_1a=Q_L-a,\quad w_2a=-Q'_L-a.
\end{equation}
To prove the reflection property for descendant operators we need to introduce auxiliary construction. This construction is an alternative free field representation for form factors or, to be more precise for the functions $J_{N,a}^{g}$ defined in~\eqref{FFD:fJ}. 


\subsection{The stripped bosonization}\label{SB}

From~\eqref{FFD:fJ} it follows that each form factor is proportional to the function $J_{N,a}^g$ up to uniform factor which depends on the number of particles only. For the exponential operators, i.e. $g=1$, the free field representation for the functions $J_{N,a}$ was proposed in~\cite{A}. Let us briefly recall the construction. Thereafter we extend this free field representation to the case of descendant operators.

Let us consider a Heisenberg algebra with the generators $d_{n}^\pm$, $n\in\mathbb{Z},\ n\neq0$. These generators satisfy the following commutation relations
\begin{equation}\label{SB:com}
	[d_m^\pm,d_n^\pm]=0,\qquad[d_m^\pm,d_n^\mp]=m A^\pm_n\delta_{m+n,0},
\end{equation}
where the coefficients $A^\pm_n$ are given by
\begin{equation}\label{SB:A}
A^\pm_n=4\,\omega^{\pm\frac32}\cos\frac{\pi n }{6}\Bigl(\cos\frac{\pi n}{3}-\cos\frac{\pi(b-b^{-1})n}{3Q}\Bigr).
\end{equation}
Note that
\begin{equation}
	A_n^-=A_{-n}^+=(-1)^n A_n^+.
\end{equation}
Let $\hat a$ be a central element of the algebra. Besides, let $|1\rangle_a$ be the vacuum state satisfying the annihilation conditions for the positive generators $d^\pm_n|1\rangle_a=0\ (n>0)$ such that $\hat a |1\rangle_a=a|1\rangle_a$. The Fock space generated by negative elements $d_{-n}^\pm\ (n>0)$ from the vacuum $|1\rangle_a$ will be denoted by $\mathcal D_a^L$. Similarly, we introduce the Fock space $\mathcal D_{a}^R$ which is generated by positive elements $d_n^\pm\ (n>0)$ from the vacuum ${}_a\langle1|$. This vacuum satisfies annihilation conditions for negative modes and ${}_a\langle1|\hat a={}_a\langle1|a$. The Fock spaces we introduced possess decompositions into the level subspaces $\mathcal D_{a}^{R}=\oplus_{n=0}^\infty \mathcal D_{a,n}^R$ and $\mathcal D_{a}^{L}=\oplus_{n=0}^\infty \mathcal D_{a,n}^L$. 

Let us consider the exponential operators
\begin{equation}\label{SB:lambda}
	\lambda^\pm(z)=\exp\sum_{n\neq0}\frac{d_n^\pm}{n}z^{-n},\quad \lambda^0(z)=:\lambda^+(z \omega^{-1})\lambda^-(z \omega):.
\end{equation}
Here we use the notation $:\cdots:$ for the normal ordering procedure which is defined by the relations
\begin{equation}
	\begin{gathered}
	\lambda^\pm(z')\lambda^\pm(z')=:\lambda^\pm(z')\lambda^\pm(z'):,\\
	\lambda^-(z')\lambda^+(z)=\lambda^+(z)\lambda^-(z')=f\Bigl(\frac{z}{z'}\omega\Bigr)f\Bigl(\frac{z}{z'}\omega^2\Bigr):\lambda^+(z)\lambda^-(z'):,
	\end{gathered}
\end{equation}
Let us define the current 
\begin{equation}\label{SB:t}
	t(z)=e^{\i\pi p}\lambda^+(z)+e^{-\i\pi p}\lambda^-(z)+h\, \lambda^0(z),
\end{equation}
which looks much the same as Lukyanov's one~\eqref{FFE:T}. The functions $J_{N,a}$ which determine the form factors of exponential operators are given by the matrix elements of the currents $t(x_i)$, namely
\begin{equation}
	J_{N,a}(x_1,\ldots,x_N)={}_a\langle1| t(x_1)\ldots t(x_N)|1\rangle_a.
\end{equation}
In order to obtain the free field representation for the functions $J_{N,a}^g$ we need an additional construction. Let us consider two representations of the algebra $\mathcal A$ in the Heisenberg algebra, $\pi_R$ and $\pi_L$, defined as follows
\begin{equation}
	\pi_R(\alpha_{-n})=\frac{d^+_n-d^-_n}{A_n^+},\quad\pi_L(\alpha_{-n})=\frac{d_{-n}^+-d_{-n}^-}{A_n^+}\quad(n>0).
\end{equation}
It is easy to get the following commutation relations:
\begin{equation}\label{SB:comm}
	\begin{aligned}
	{}[ \pi_R(\alpha_{-n}),\lambda^\pm(z) ]& =(\pm)^{n+1}z^n \lambda^\pm(z),\\
	[ \pi_L(\alpha_{-n}),\lambda^\pm(z) ]& =-(\mp)^{n+1}z^{-n} \lambda^\pm(z),\\
	[\pi_R(\alpha_{-m}),\pi_L(\alpha_{-n})]&=-m(A_m^+)^{-1}(1+(-1)^m)\delta_{m-n,0}
	\end{aligned}
\end{equation}
We shall also use the notation
\begin{equation}\label{SB:h}
	{}_a\langle h|={}_a\langle1|\pi_R(h),\quad |\bar h\rangle_a=\pi_L(h)|1\rangle_a.
\end{equation}
Consider the chiral element $h\in\mathcal A$. Taking into account the commutation relations~\eqref{SB:comm} we easily obtain that the $J$ functions of chiral and antichiral descendant operators are given by the following matrix elements
\begin{equation}\label{SB:J}
	J^h_{N,a}(x_1,\ldots,x_N)={}_a\langle h|t_1(x_1)\ldots t_N(x_N)|1\rangle_a,\quad J^{\bar h}_{N,a}(x_1,\ldots,x_N)=\langle 1|t_1(x_1)\ldots t_N(x_N)|\bar h\rangle_a.
\end{equation}
Now, let us consider the generic element $g=h\bar h$. We denote the corresponding matrix element as follows
\begin{equation}\label{SB:Jtilde}
	\tilde J^{h\bar h'}_{N,a}(x_1,\ldots,x_N)={}_a\langle h|t_1(x_1),\ldots,t_N(x_N)|\bar h'\rangle_a.
\end{equation}
Notice that this matrix element do not coincide with the function $J_{N,a}^{h\bar h}(X)$. However, these functions can be related to each other. Indeed, by pushing the element $\pi_R(h)$ and $\pi_L(h)$ from the definition of the vectors ${}_a\langle h|$ and $|h\bar'\rangle_a$ in the expression~\eqref{SB:Jtilde} in the opposite directions we get that this function is given by certain linear combination of the functions $J_{N,a}^{h \bar h'}$, i.e.
\begin{equation}\label{SB:JtildeJ}
	\tilde J_{N,a}^{h\bar h'}(x_1,\ldots x_N)=\sum_{i}J_{N,a}^{h_i\bar h'_i}(x_1,\ldots,x_N).
\end{equation}
If in the l.h.s.\ of this expression we consider the elements from the level subspaces $h\in \mathcal A_n$ and $h'\in\mathcal A_{\bar n}$ then in the r.h.s\ the sum is taken over the $J$ functions corresponding to the elements $h_i\in\mathcal A_{n_i}$ and $h'_i\in\mathcal A_{\bar n'_i}$ such that $n_i\leq n$ and $\bar n_i\leq \bar n$.

The vectors ${}_a\langle h|$ and $|\bar h'\rangle_a$ will be referred to as the `physical' ones since their matrix elements can be expressed as a linear combination of $J$ functions. These vectors form a subspaces in the spaces $\mathcal D_a^R$ and $\mathcal D_a^L$. The `physical' subspaces can be decomposed into the direct sum over the level subspaces $\mathcal D_{a}^{R,{\rm phys}}=\oplus_{n=0}^\infty \mathcal D_{a,n}^{R,{\rm phys}}$ and $\mathcal D_{a}^{L,{\rm phys}}=\oplus_{n=0}^\infty \mathcal D_{a,n}^{L,{\rm phys}}$. Evidently that
\begin{equation}
	\dim \mathcal D_{a,n}^{R,{\rm phys}}=\dim \mathcal D_{a,n}^{L,{\rm phys}}=\dim \mathcal F_n.
\end{equation}
Further we shall use an alternative definition of the `physical' subspaces. Namely, these subspaces can be defined as kernels of the operators $D_{n}$ defined as follows
\begin{equation}\label{SB:D}
	D_n=\omega^{\frac{3}{2}n} d^+_n+\omega^{-\frac{3}{2}n} d^-_n,\quad (n\neq0).
\end{equation}
It is easy to check the consistency of these two definitions. Indeed, the operators $D_n$ satisfy the following commutation relations
\begin{equation}
[D_n,\pi_R(h)]=0,\quad[D_n,\pi_L(h)]=0,\quad[D_n,D_m]=0.
\end{equation}
Besides, one can show that for any state ${}_a\langle v|\in \mathcal D_a^R$  such that ${}_a\langle v|D_{-n},\ n>0$, there exist the element $h\in\mathcal A$ such that ${}_a\langle v|={}_a\langle1|\pi_R(h)$. Similarly, for any state $| v\rangle_a\in\mathcal D_{a}^L$ such that $D_{n}|v\rangle_a=0,\ n>0$, there exist the element $h\in\mathcal A$ such that $| v\rangle_a=\pi_L(h)|1\rangle_a$.
%


\subsection{Analytic properties of the $J_{N,a}^g$ functions}

The form factors of exponential operators are known to satisfy recurrence relations~\cite{A}. One can obtain the similar relations for the form factors of descendant operators. Indeed, the function $J^g_{N+1,a}(z,X)$ is an analytic function in the variable $z$ depending on the parameters $X=\lbrace x_1,\ldots,x_N\rbrace$. This function has simple kinematical and bound state poles at points prescribed by the form factor axioms. The residues at these poles can be evaluated explicitly as shown in~\cite{A}. One can separate the kinematical and the bound state poles contribution from the regular part and obtain
\begin{multline}\label{FFD:RecInf}
		J^g_{N+1,a}(z,X)=J^{(\infty)g}_{N+1,a}(z,X)+\sum\limits_{n=1}^N\frac{x_n}{z+x_n}K_n(X)J^g_{N-1,a}(\hat X_n)+\\
		+\sum\limits_{n=1}^N\frac{x_n\omega^{2}}{z-x_n\omega^2}B_n^+(X)J^g_{N,a}(x_n\omega,\hat X_n)-\sum_{n=1}^N\frac{x_n\omega^{-2}}{z-x_n\omega^{-2}}B_n^-(X)J^g_{N,a}(x_n\omega^{-1},\hat X_n),
\end{multline}
Here we introduce the notation $\hat X_n=X\setminus \lbrace x_n\rbrace$ and the functions $K_n(X)$ and $B_n^\pm(X)$ are given by
\begin{equation}\label{FFD:BK}
	\begin{aligned}
		K_n(X)&=-\i\frac{(h^2-3)(h^2-1)}{2\sqrt3}\biggl(\prod_{i\neq n}f\Bigl(\frac{x_n}{x_i}\omega^{2}\Bigr)f\Bigl(\frac{x_n}{x_i}\omega\Bigr)-\prod_{i\neq n}f\Bigl(\frac{x_n}{x_i}\omega^{-2}\Bigr)f\Bigl(\frac{x_n}{x_i}\omega^{-1}\Bigr)\biggr),\\
		B_n^{\pm}(X)&=-\i\frac{ h ( h ^2-1)}{\sqrt3}\prod_{i\neq n} f\Bigl( \frac{x_n}{x_i}\omega^{\pm1} \Bigr).
	\end{aligned}
\end{equation}

The derivation of the expansion~\eqref{FFD:RecInf} repeats the main features of those one for the exponential operators and we omit it here. The regular part is given by the function $J^{(\infty)g}_{N+1,a}(z,X)$. This function is regular everywhere except the points $z=0$ and $z=\infty$. Since the sum over residues is of the order $O(z^{-1})$ as $z\to\infty$ the asymptotic behavior of the $J^g$ function as a function of $z$ is governed by the regular part $J^{(\infty)g}$. Further we shall also use another expansion of the $J^g$ function in the vicinity of $z=0$ which is given by
\begin{multline}\label{FFD:Rec0}
		J^g_{N+1,a}(z,X)=J^{(0)g}_{N+1,a}(z,X)-\sum\limits_{n=1}^N\frac{x^{-1}_n}{z^{-1}+x_n^{-1}}K_n(X)J^g_{N-1,a}(\hat X_n)-\\
		-\sum\limits_{n=1}^N\frac{x_n^{-1}\omega^{-2}}{z^{-1}-x_n^{-1}\omega^{-2}}B_n^+(X)J^g_{N,a}(x_n\omega,\hat X_n)+\sum_{n=1}^N\frac{x_n^{-1}\omega^{2}}{z^{-1}-x_n^{-1}\omega^{2}}B_n^-(X)J^g_{N,a}(x_n\omega^{-1},\hat X_n).
\end{multline}
In this case the asymptotic behavior is governed by the function $J^{(0)g}_{N+1,a}(z,X)$. From expressions~\eqref{FFD:RecInf} and~\eqref{FFD:Rec0} we easily obtain that the regular parts of these expansions are related as follows
\begin{equation}\label{FFD:D}
	J_{N+1,a}^{(0)g}(z,X)-J_{N+1,a}^{(\infty)g}(z,X)=D^g_{N,a}(X),
\end{equation}
where the function $D^g_{N,a}(X)$ is given by
\begin{multline}
	D_{N,a}^g(X)=\sum_{n=1}^N x_n \omega^2 B^+_n(X)J^g_{N,a}(x_n \omega;\hat X)-\sum_{n=1}^N x_n \omega^{-2}B^-_n(X)J^g_{N,a}(x_n \omega^{-1};\hat X_n)+\\
	+\sum_{n=1}^N K_n(X)J_{N-1,a}^g(\hat X_n).
\end{multline}

The expansions~\eqref{FFD:RecInf} and~\eqref{FFD:Rec0} allows us to calculate form factors recursively. The recursion relations for exponential operators, i.e.\ $g=1$, immediately follows from these expansion since the regular parts $J^{(0)}$ and $J^{(\infty)}$ can be easily calculated~\cite{A}. In Section~\ref{E} we use these expansions to obtain recurrence relations for the chiral level $2$ descendant operators.


\subsection{Reflection relations for descendant operators}

In this subsection we shall prove that the form factors of descendant operators satisfy the reflection relations. More precisely, we prove this statement for the matrix elements~\eqref{SB:Jtilde}. However, from~\eqref{SB:JtildeJ} it follows that if the reflection relations hold for $\tilde J$ functions these relations hold for $J$ functions as well.
\begin{theorem}\label{RR:th}
For generic values of the parameter $a$ there exist a representation $r_a(w)$ of the group $\mathcal W$ on the algebra $\mathcal A$ such that for any elements $h,h'\in\mathcal A$ the following relation holds
\begin{equation}\label{RR:JJ}
	\tilde J_{N,a}^{h\bar h'}(x_1,\ldots,x_N)=\tilde J_{N,wa}^{(r_a(w)h)(\overline{r_{-a}(w)h'})}(x_1,\ldots,x_N).
\end{equation}
\end{theorem}
The proofs of this theorem is very similar to those of~\cite{FL}. The main idea is that all form factors can be obtained as a coefficients of the large rapidity expansion of the form factors of the exponential operators~\cite{FPP},~\cite{FP}.

First, let us introduce auxiliary current
\begin{equation}\nonumber
	s(z)=:\lambda^+(z \omega^{-3/2})\lambda^-(z \omega^{3/2}):,
\end{equation}
such that
\begin{equation}\nonumber
	s(z)t(x)=t(x)s(z)=f\Bigl(\frac{z}{x}\omega^{1/2}\Bigr)f\Bigl(\frac{z}{x}\omega^{-1/2}\Bigr):s(z)t(x):.
\end{equation}
It is easy to check that the matrix element of some number of the currents $t(x_i)$ and $s(y_i)$ satisfy the reflections relations, namely
\begin{equation}\nonumber
	{}_a\langle1|t(x_1)\ldots t(x_K)s(y_1)\ldots s(y_L) |1\rangle_a={}_{wa}\langle1|t(x_1)\ldots t(x_K)s(y_1)\ldots s(y_L) |1\rangle_{wa}.
\end{equation}
Indeed, the matrix elements of the currents $t(x_i)$ satisfy the reflection property while an insertion of the currents $s(y_i)$ in matrix elements results in the $a$-independent overall factor.

Now let us prove the reflection property. First, we prove that the product of the currents $t(x_i)$ and $s(y_i)$ acting on any highest weight vector spans the whole Fock module corresponding to this vector. Consider the expansion
\begin{equation}\nonumber
{}_a\langle 1|t(\xi_1^{-1}z)\ldots t(\xi_p^{-1}z)s(\eta_1^{-1}z)\ldots s(\eta_q^{-1}z)=\sum_{n=0}^\infty z^{-n}\, {}_a\langle n;\xi_1,\ldots,\xi_p;\eta_1,\ldots,\eta_q|.
\end{equation}
Hereinafter we shall use the notation $\Xi=(\xi_1,\ldots,\xi_p)$ and ${\rm H}=(\eta_1,\ldots,\eta_q)$. We propose that for generic values of the parameter $a$ and large enough values of $p$ and $q$ it is possible to choose a set $\Xi^{(i)}$ and ${\rm H}^{(i)}$ with $i=1,\ldots\dim(\mathcal F^{\otimes2})_n$ such that the vectors ${}_a\langle n;\Xi^{(i)},{\rm H}^{(i)}|$ form a basis in the level subspace $\mathcal D_{a,n}^R$. Let us prove this proposition in the limit $a\to -i\infty$. In this limit we have
\begin{multline}\nonumber
	{}_a\langle 1|t(\xi_1^{-1}z)\ldots t(\xi_p^{-1}z)s(\eta_1^{-1}z)\ldots s(\eta_q^{-1}z)=\\ = F(\xi_1,\ldots,\xi_p,\eta_1,\ldots,\eta_q){}_a\langle 1|:\lambda^+(\xi_1^{-1}z)\ldots \lambda^+(\xi_p^{-1}z)s(\eta_1^{-1}z)\ldots s(\eta_q^{-1}z):.
\end{multline}
Here $F(\xi_1,\ldots,\xi_p,\eta_1,\ldots,\eta_q)$ is some irrelevant for our purpose function and the normal ordered product in the r.h.s. is of the form
\begin{equation}\nonumber
	:\lambda^+(\xi_1^{-1}z)\ldots \lambda^+(\xi_p^{-1}z)s(\eta_1^{-1}z)\ldots s(\eta_q^{-1}z):=:\exp\left(\sum_{n\neq0}\frac{\kappa^+_n d^+_n+\kappa^-_n d^-_n}{n}z^{-n}\right):,
\end{equation}
where
\begin{equation}\nonumber
	\kappa^+_n=\sum_{i=1}^p\xi^n+\omega^{\frac{3}{2}n}\sum_{j=1}^q \eta_j^n,\quad \kappa^-_n=\omega^{-\frac{3}{2}n}\sum_{j=1}^q\eta_j^n.
\end{equation}
Let us consider the following expansion
\begin{equation}\nonumber
{}_a\langle1|:\lambda^+(\xi_1^{-1}z)\ldots \lambda^+(\xi_p^{-1}z)s(\eta_1^{-1}z)\ldots s(\eta_q^{-1}z):=\sum_{n=0}^\infty z^{-n}\cdot{}_{(-)}\langle n;\Xi;{\rm H}|.
\end{equation}
Then we get
\begin{equation}\nonumber
	{}_{(-)}\langle n;\Xi;{\rm H}|={}_a\langle1|\sum_{s=1}^n\sum_{\substack{n_1,\ldots,n_s>0\\ n_1+\ldots+ n_s=n}}C_{n_1\ldots n_s}\prod_{i=1}^s(\kappa^+_{n_i}d^+_{n_i}+\kappa^-_{n_i}d^-_{n_i})
\end{equation}
with some nonzero constants $C_{n_1\ldots n_s}$. Notice that all possible products of $d_n^\pm$ enter the r.h.s.

For large enough values of $p$ and $q$ the functions $\kappa^\pm_i$ with $1\leq i\leq n$ are functionally independent and can be considered as an independent variables. Besides, the monomials $\kappa_{\sigma_1}^\pm\ldots\kappa_{n_s}^\pm$ are linearly independent. Consequently, for any set of the numbers $A_{n_1\ldots n_s}^{\sigma_1\ldots \sigma_s}$ with $s=1,\ldots,n$ and $n_1,\ldots, n_s>0$, $n_1+\ldots +n_s=n$, we have
\begin{equation}\nonumber
	\sum_s\sum_{\substack{\sigma_1,\ldots,\sigma_n\\n_1,\ldots n_s}}\overline{A_{n_1\ldots n_s}^{\sigma_1\ldots \sigma_s}}\kappa_{n_1}^{\sigma_1}\ldots \kappa_{n_s}^{\sigma_s}\neq0
\end{equation}
for some values of $\kappa_1^\pm,\ldots\kappa_n^\pm$. Hence, the vector generated by the numbers $A_{n_1\ldots n_s}^{\sigma_1\ldots \sigma_s}$ is not orthogonal to at least one vector generated by the product of $\kappa_m^\pm$. It means that there is no vector in the $\dim (\mathcal F^{\otimes2})_n$-dimensional space orthogonal to all vectors generated by the product of $\kappa_{m}^\pm$ for any values of $\Xi$ and ${\rm H}$. Consequently, the vectors ${}_{(-)}\langle n;\Xi;{\rm H}|$ for some values $\Xi^{(i)}$ and ${\rm H}^{(i)}$ with $i=1,\ldots \dim(\mathcal F^{\otimes 2})_n$ form a basis in the level $n$ subspace of the Fock module. The deformation argument proves the same statement for the generic values of the parameter $a$. Hence we prove that the product of the currents $t(x_i)$ and $s(y_i)$ acting on any highest weight vector spans the whole Fock module corresponding to this vector.

Now let us prove that the reflections act on the space of the matrix elements with respect to the whole Fock module.  Let ${}_a\langle n;i|={}_a\langle n;\Xi^{(i)};{\rm H}^{(i)}|$, $i=1,\ldots \dim(\mathcal F^{\otimes2})_n$ be a basis vectors in the level $n$ subspace of the right Fock module. Similarly, let the states $|\bar n;j\rangle_a$ be a basis vectors in the level $\bar n$ subspace of the left Fock module. As follows from the reflection properties for exponential operators, we have
\begin{equation}\nonumber
	{}_a\langle n;i|t(x_1)\ldots t(x_N)|\bar n;j\rangle_a={}_{wa}\langle n;i|t(x_1)\ldots t(x_N)|\bar n;j\rangle_{wa}.
\end{equation}
Let us prove that the restriction of the whole Fock module to the `physical' one preserves the reflection property. Let the vectors ${}_a\langle 1|\pi_R(h_{a,n,\mu})={}_a\langle\widetilde{n,\mu}|=\sum_i v_i^\mu(a){}_a\langle n;i|$ form a basis in the $\mathcal D_{a,n}^{R,{\rm phys}}$. Similarly, let $\pi_L(h'_{a,\bar n,\nu})|1\rangle_a=|\widetilde{\bar n;\nu}\rangle_a=\sum_j \bar v_j^\nu(a)|\bar n,j\rangle_a$ be a basis in the $\mathcal D_{a,n}^{L,{\rm phys}}$. We define the `physical' states by means of the operator $D_m$ introduced in~\eqref{SB:D}. Notice that the auxiliary current $s(z)$ is given by the following expression
\begin{equation}\nonumber
	s(z)=\exp\biggl(\sum_{m\neq0}\frac{D_m }{m}z^{-m}\biggr).
\end{equation}
From this relation we easily get
\begin{multline}\nonumber
	{}_a\langle1|t(x_1)\ldots t(x_M)s(y_1)\ldots s(y_N) D_m t(x'_1)\ldots t(x'_{M'})s(y'_1)\ldots s(y_{N'})|1\rangle_a=\\
	{}_{wa}\langle1|t(x_1)\ldots t(x_M)s(y_1)\ldots s(y_N) D_m t(x'_1)\ldots t(x'_{M'})s(y'_1)\ldots s(y_{N'})|1\rangle_{wa}.
\end{multline}
We have
\begin{equation}\nonumber
	0={}_a\langle\widetilde{n;\mu}|D_{-m}|n-m;j\rangle_a=\sum_i v_i^\mu(a){}_a\langle n;i|D_{-m}|n-m;j\rangle_a=\sum_i v_i^\mu(a){}_{wa}\langle n;i|D_{-m}|n-m;j\rangle_{wa}.
\end{equation}
Therefore,
\begin{equation}\nonumber
	\sum_i v_i^\mu(a){}_{wa}\langle n;i|D_{-m}=0.
\end{equation}
From this equation we immediately conclude that there exist an element $h_{wa,n\mu}^w$ such that
\begin{equation}\nonumber
	{}_{wa}\langle1|\pi_R(h^w_{wa,n,\mu})=\sum_i v_i^\mu(a){}_{wa}\langle n;i|.
\end{equation}
Similarly, there exist an element $h'^w_{wa,n,\mu}$ such that
\begin{equation}\nonumber
	\pi_L(h'^w_{wa,n,\mu})|1\rangle_{wa}=\sum_j \bar v_j^\nu(a)|n;j\rangle_{wa}.
\end{equation}
Finally, we get
\begin{multline}\nonumber
	{}_a\langle1|\pi_R(h_{a,n,\mu})t(x_1)\ldots t(x_N)\pi_L(h'_{a,\bar n,\nu})|1\rangle_a={}_a\langle\widetilde{n;\mu}|t(x_1)\ldots t(x_N)|\widetilde{\bar n,\nu}\rangle_a=\\
=\sum_{i,j}v_i^\mu(a)\bar v_j^\nu(a){}_a\langle n;i|t(x_1)\ldots t(x_N)|\bar n;j\rangle_a=\sum_{i,j}v_i^\mu(a)\bar v_j^\nu(a){}_{wa}\langle n;i|t(x_1)\ldots t(x_N)|\bar n;j\rangle_{wa}=\\
	={}_{wa}\langle1|\pi_R(h_{wa,n,\mu}^w)t(x_1)\ldots t(x_N)\pi_L(h'^w_{wa,\bar n,\nu})|1\rangle_{wa}.
\end{multline}
We obtain the map $r_a(w)$ on the subspaces $\mathcal D_a^{R,{\rm phys}}$ and $\mathcal D_a^{L,{\rm phys}}$, namely
\begin{equation}\nonumber
	r_a(w)({}_a\langle1|\pi_R(h_{a,n,\mu}))={}_{wa}\langle1|\pi_R(h^w_{wa,n,\mu}),\quad r_a(w)(\pi_L(h'_{a,\bar n,\nu})|1\rangle_a)=\pi_L(h'^w_{wa,\\bar n,\nu})|1\rangle_{wa}.
\end{equation}
Comparing with the property we  conclude that
\begin{equation}\nonumber
	r_a(w)h_{a,n,\mu}=h^w_{wa,n,\mu},\quad r_{-a}(w)h'_{a,\bar n,\nu}=h'^w_{wa,\bar n,\nu}.
\end{equation}
This proves Theorem~\ref{RR:th}.


\section{Explicit example: level $(2,0)$ descendants}\label{E}

In this section we obtain recurrence relations for the form factors of descendant operators at the chiral level $(2,0)$. Taking into account expansion~\eqref{FFD:RecInf} it is sufficient to find the regular part of the function $J^g_{N,a}(z,X)$. This means that we need to consider the asymptotic of this function as $z\to\infty$.

The level $(2,0)$ subspace is spanned on the elements $\alpha_{-2}$ and $\alpha_{-1}^2$ acting on the highest weight vector~${}_a\langle1|$. For these elements we get the following expansion
\begin{equation}\label{RR:JInf}
	J_{N+1,a}^g(z,X)=C^{(\infty)g}_{N,a,2}(X)\cdot z^2+C^{(\infty)g}_{N,a,1}(X)\cdot z+C^{(\infty)g}_{N,a,0}(X)+O(z^{-1}),
\end{equation}
where the coefficients are given by
\begin{equation}\nonumber
	\begin{aligned}
	C^{(\infty)\alpha_{-1}^2}_{N,a,2}(X)&=J_{1,a}\cdot J_{N,a}(X),\\
	C^{(\infty)\alpha_{-1}^2}_{N,a,1}(X)&=\frac{h^2-1}{2}(2\sqrt3\sin\pi p+h)J_{N,a}^{\alpha_{-1}}+\frac{\i\sqrt3(h^2-1)}{2}J_{1,a}\cdot L_{N,a}(X),\\
	C^{(\infty)\alpha_{-2}}_{N,a,2}(X)&=J_{1,a}^{\alpha_{-2}}\cdot J_{N,a}(X),\\
	C^{(\infty)\alpha_{-2}}_{N,a,1}(X)&=-\frac{\i\sqrt3(h^2-1)}{2}J_{1,a}J_{N,a}^{\alpha_{-1}}+\frac{\i\sqrt3(h^2-1)}{2}J_{1,a}^{\alpha_{-2}}\cdot L_{N,a}(X),\\
	\end{aligned}
\end{equation}
and
\begin{multline}\nonumber
	L_{N,a}(X)=\sum_{I_++I_-+I_0=I}h^{\#I_0}e^{(\#I_+-\#I_-)\i\pi p}\Bigl(\sum_{i\in I_+}x_i-\sum_{j\in I_-}x_j-\frac{\i}{\sqrt3}\sum_{k\in I_0}x_k\Bigr)\times\\ 
	\times\prod_{i\in I_+,j\in I_-,k\in I_0}f\Bigl(\frac{x_i}{x_j}\omega\Bigr)f\Bigl(\frac{x_i}{x_j}\omega^2\Bigr)f\Bigl(\frac{x_i}{x_k}\omega\Bigr)f\Bigl(\frac{x_j}{x_k}\omega^{-1}\Bigr)\prod_{(p<q)\in I_0}f\Bigl(\frac{x_p}{x_q}\Bigr).
\end{multline}
For the recurrence relations we also need the coefficients $C_{a,N,0}^{(\infty)g}(X)$ in the expansions~\eqref{RR:JInf}. It is rather complicated to find them directly. However, it is not necessary since these coefficients are related with the leading coefficients in another limit $z\to0$. The expansion of the $J^g_{N,a}$ function in the vicinity of this point is given by
\begin{equation}\nonumber
	J_{N+1,a}^g(z,X)=C^{(0)g}_{N,a,0}(X) +C^{(0)g}_{N,a,1}(X)\cdot z+C^{(0)g}_{N,a,2}(X)\cdot z^2+O(z^{3}).
\end{equation}
By using factorization property~\eqref{FP:f} we immediately get the coefficients $C^{(0)g}_{a,N,0}(X)$, namely
\begin{equation}\nonumber
	C_{N,a,0}^{(0)\alpha_{-1}^2}=J_{1,a}\cdot J_{N,a}^{\alpha_{-1}^2}(X),\quad
	C_{N,a,0}^{(0)\alpha_{-2}}=J_{1,a}\cdot J_{N,a}^{\alpha_{-2}}(X).
\end{equation}
As immediately follows from~\eqref{FFD:D} the required coefficients are given by
\begin{equation}\nonumber
	C_{N,a,0}^{(\infty)g}=C_{N,a,0}^{(0)g}-D_{N,a}.
\end{equation}
Consequently, expansion~\eqref{FFD:Rec0} becomes a recurrence relation for the form factors of descendant operators at the level $(2,0)$ where the regular part is of the form
\begin{equation}\label{RR:Jreg}
	J^{(0)g}_{N+1,a}(z,X)=C_{N,a,2}^{(\infty)g}(X)\cdot z^2+C_{N,a,1}^{(\infty)g}(X)\cdot z+C_{N,a,0}^{(0)g}(X).
\end{equation}
Now let us find such combinations of the elements $\alpha_{-2}$ and $\alpha_{-1}^2$ which are invariant with respect to the reflections. It is easy to check that the required combinations are given by
\begin{equation}\nonumber
	\begin{aligned}
	h_a^{(11)}&=\alpha_{-1}^2,\\
	h_a^{(2)}&=\frac{1}{(h^2-1)(\sqrt3+2h\cos\pi(p+1/6))}\Bigl(\alpha_{-2}-\frac{J^{\alpha_{-2}}_{1,a}}{J_{1,a}}\alpha_{-1}^2\Bigr).
	\end{aligned}
\end{equation}
Indeed, the element $\alpha_{-1}$ corresponds to the integral of motion~\eqref{IM:f}, consequently
\begin{equation}\nonumber
	J_{N,a}^{h_a^{(11)}}(x_1,\ldots,x_N)=(x_1+\ldots+x_N)^2\cdot J_{N,a}(x_1,\ldots,x_N),
\end{equation}
and the functions $J_{N,a}$ are known to be invariant with respect to reflections.

Now let us consider consider the element $h_a^{(2)}$. We can easily obtain recurrence relation for this element. This relation is given by the expansion~\eqref{FFD:Rec0} where the regular part is given by the appropriate linear combination of the functions~\eqref{RR:Jreg}, namely
\begin{equation}\label{RR:Jh2}
	J_{N+1,a}^{(0)h_a^{(2)}}(z,X)=-z\frac{2\i}{J_{1,a}}J_{N,a}^{\alpha_{-1}}(X)+J_{1,a}\cdot J_{N,a}^{h_a^{(2)}}(X).
\end{equation}
This relation together with the initial conditions
\begin{equation}\nonumber
	J_{0,a}^{h_a^{(2)}}=J_{1,a}^{h_a^{(2)}}(x)=0,\quad
	J_{1,a}=2\cos\pi p+h
\end{equation} defines the set of functions corresponding to the element $h_a^{(2)}$ uniquely. The invariance of this set of functions with respect to the reflections generated by the group $\mathcal W$ simply follows from~\eqref{RR:Jh2}. Indeed, one can easily check that the parameter $a$ always enters the recurrence relations by means of the function $\cos\pi p$ which is invariant with respect to reflections.

The recurrence relations allows us to calculate the first few $J$ functions at the level $(2,0)$ easily, for example
\begin{equation}\nonumber
	\begin{aligned}
	J_{2,a}^{h_a^{(2)}}(x_1,x_2)&=-2\i\,x_1x_2,\\
	J_{3,a}^{h_a^{(2)}}(x_1,x_2,x_3)&=2\i\,\frac{J_{1,a}\cdot \sigma_2^4+(J_{1,a}-h^3+h)\sigma_1^3\sigma_2\sigma_3-J_{1,a}\cdot \sigma_1^2\sigma_2^3-h(4-5h^2+h^4)\sigma_1^2\sigma_3^2}{\prod_{i<j}(x_i^2+x_i x_j +x_j^2)},
	\end{aligned}
\end{equation}
where we introduce the set of elementary symmetric polynomials $\sigma_i(x_1,\ldots,x_N)$ defined by the generating function
\begin{equation}\nonumber
	\prod_{i=1}^N(x+x_i)=\sum_{i=0}^N x^{N-i}\sigma_i(x_1,\ldots,x_N).
\end{equation}
In principle, one can obtain recurrence relations for the form factors of descendant operators at each level. However, an explicit expression of these relations is rather cumbersome and not too useful. The recurrence relations provides a convenient framework to study analytical properties of form factor.

\section{Conclusion}\label{C}

In this paper we use the free field approach to construct the space of solutions to the form factors axioms. We prove that there exist a bijection between this space and the Fock space of descendant operators for a generic values of the parameter $a$. The proposed free field representation provides a convenient framework to study analytical properties of the form factors. However, the problem of identification between the form factors and particular descendant operators is not solved. The cluster factorization property and the correct asymptotic behavior do not solve the identification problem except for the case of exponential operators. The reflection relations between the form factors might be a step forward to the solution of this problem. The further speculations about the identification problem involve the resonant identities between the operators~\cite{ML}.

The quantum group restrictions of the Bullough-Dodd model are known to describe certain classes of perturbed minimal models. It is natural to conjecture that the lightest breather form factors for any local operators in the perturbed minimal models can be analytically continued from the Bullough-Dodd case. We checked this conjecture for the form factors of primary operators in the Ising model in a magnetic field~\cite{A}. Therefore, the proposed construction provides a convenient framework to study form factors of local operators in certain classes of the perturbed minimal models.


\section*{Acknowledgments}
I am grateful to M.~Lashkevich and Ya.~Pugai for encourage discussions. I am also
grateful to C.~Rim for his hospitality at the Sogang University. The work was supported by the Russian Ministery of Education 2012-1.1-12-000-1011-016, RFBR research project No. 12-02-33011 and the National Research Foundation of Korea (NRF) grant funded by the Korea government (MEST) through the Center for Quantum Spacetime (CQUeST) of Sogang University with grant number 2005-0049409.

\end{document}